\pdfoutput=1

\documentclass[11pt]{article}

\usepackage[final]{acl}

\usepackage{times}
\usepackage{latexsym}

\usepackage[T1]{fontenc}

\usepackage[utf8]{inputenc}

\usepackage{microtype}

\usepackage{inconsolata}

\usepackage{graphicx}

\usepackage{hyperref}       
\usepackage{url}            
\usepackage{latexsym}
\usepackage{multirow}
\usepackage{subfigure}
\usepackage{amsmath}
\usepackage{booktabs}       
\usepackage{amssymb}
\usepackage{microtype}      
\usepackage{balance}
\usepackage{inconsolata}
\usepackage{adjustbox}
\usepackage{graphicx}
\usepackage[utf8]{inputenc} 
\usepackage[T1]{fontenc}    
\usepackage{amsfonts}       
\usepackage{nicefrac}       
\usepackage{xcolor}         
\usepackage[english]{babel}
\usepackage{moresize}
\usepackage{algorithmic}
\usepackage{comment}
\usepackage{paralist}
\usepackage{bm}
\usepackage{pgfplots}
\usetikzlibrary{pgfplots.dateplot}
\usepackage{flushend}
\usepackage{bbm}
\usepackage{longtable}
\usepackage{rotating}
\usepackage{mathrsfs}
\usepackage{enumitem}
\usepackage[linesnumbered,algoruled,boxed,lined]{algorithm2e}
\usepackage{filecontents}
\usepackage{wrapfig}
\usepackage{caption}
\usepackage{bbding}
\usepackage{float}
\usepackage[utf8]{inputenc}
\usepackage[most]{tcolorbox} 
\usepackage{amsmath}
\usepackage{listings} 

\newtcolorbox[auto counter, number within=section]{blue-box}[2][]{colback=blue!10!white, colframe=blue!75!black, coltitle=white, fonttitle=\bfseries, title={#2}, 
  boxrule=0.5mm, left=0mm, right=0mm, top=0mm, bottom=0mm, #1}
  
\makeatletter
\def\blfootnote{\gdef\@thefnmark{}\@footnotetext}
\makeatother

\title{Aria-UI: Visual Grounding for GUI Instructions}

\author{
 \textbf{Yuhao Yang\textsuperscript{1}},
 \textbf{Yue Wang\textsuperscript{3}},
 \textbf{Dongxu Li\textsuperscript{4}},
 \textbf{Ziyang Luo\textsuperscript{2}},
\\
 \textbf{Bei Chen\textsuperscript{5}},
 \textbf{Chao Huang$^{\dagger}$\textsuperscript{1}},
 \textbf{Junnan Li$^{\dagger}$\textsuperscript{2}}
\\
\\
 \textsuperscript{1}The University of Hong Kong,
 \textsuperscript{2}Salesforce AI Research,
 \textsuperscript{3}Alibaba Group,
 \\
 \textsuperscript{4}Australian National University,
 \textsuperscript{5}Independent Researcher
 \\
 \\
 \centering{ \href{https://ariaui.github.io}{\texttt{https://ariaui.github.io}}}
}

\def\model{Aria-UI\xspace}
\def\modelth{Aria-UI$_{TH}$\xspace}
\def\modelih{Aria-UI$_{IH}$\xspace}
\newcommand{\paratitle}[1]{\noindent\textbf{#1}}

\begin{document}
\maketitle

\blfootnote{\textdagger~Corresponding author(s).}

\begin{abstract}
Digital agents for automating tasks across different platforms by directly manipulating the GUIs are increasingly important. For these agents, grounding from language instructions to target elements remains a significant challenge due to reliance on HTML or AXTree inputs. In this paper, we introduce \model, a large multimodal model specifically designed for GUI grounding. \model adopts a pure-vision approach, eschewing reliance on auxiliary inputs. To adapt to heterogeneous planning instructions, we propose a scalable data pipeline that synthesizes diverse and high-quality instruction samples for grounding. To handle dynamic contexts in task performing, \model incorporates textual and text-image interleaved action histories, enabling robust context-aware reasoning for grounding. \model sets new state-of-the-art results across offline and online agent benchmarks, outperforming both vision-only and AXTree-reliant baselines. We release all training data and model checkpoints at \url{https://ariaui.github.io} to foster further research.
\end{abstract}

\section{Introduction}
\begin{figure*}[h]
    \includegraphics[trim=5mm 0 5mm 0, clip,width=\linewidth]{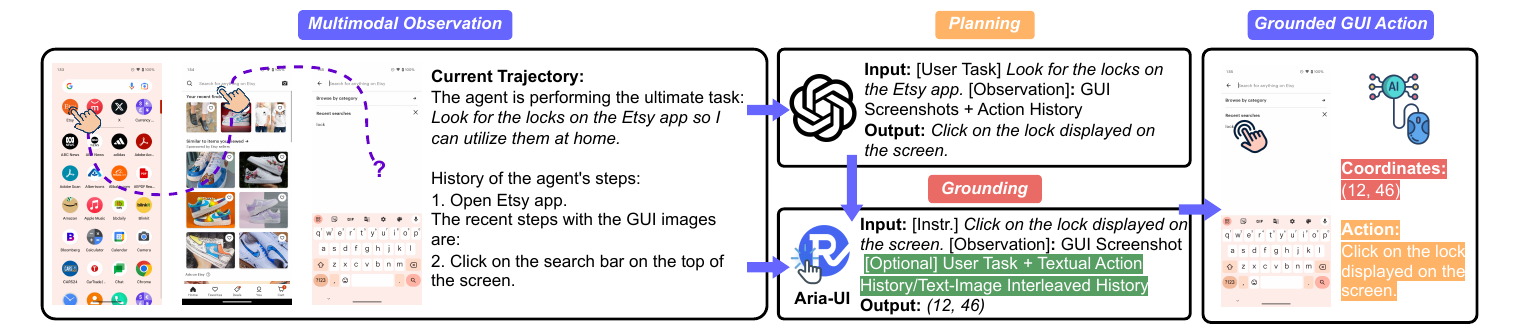}
    \vspace{-0.3in}
    \caption{The two-stage task performing process for general GUI agents. \model serves as a robust grounding model to make the planned actions truly happen.}
    \label{fig:overview}
    \vspace{-0.1in}
\end{figure*}

\begin{table*}[htb]
    \centering
    \resizebox{\linewidth}{!}{
\begin{tabular}{lccccccccc}
    \toprule
    \textbf{Collection} & \textbf{\#Web Img.} & \textbf{\#Mobile Img.} & \textbf{\#Desktop Img.} & \textbf{Input Text} & \textbf{Supervision} & \textbf{Open Source} & \textbf{Action History} & \textbf{\#Elements} & \textbf{\#Samples} \\
    \midrule
    Ferret-UI-AMP & / & 84K & / & Human Ann. & Point Coordinates & \textcolor{red}{\XSolidBrush} & \textcolor{red}{\XSolidBrush}  & - & 160K \\
    CogAgent-CCS400K & 400K & / & / & HTML Text & Point Coordinates & \textcolor{red}{\XSolidBrush} & \textcolor{red}{\XSolidBrush}  & 70M & - \\
    UGround-Web-Hybrid & 773K & / & / & HTML Attr. + Refer. Caption & Point Coordinates & \textcolor{red}{\XSolidBrush} & \textcolor{red}{\XSolidBrush} & 18.1M & 9M \\
    UGround-Web-Direct & 408K & / & / & Refer. Caption & Point Coordinates & \textcolor{red}{\XSolidBrush} & \textcolor{red}{\XSolidBrush}  & 408K & 408K \\
    SeeClick & 270K & / & / & HTML Text & Point Coordinates & \textcolor{green}{\Checkmark} & \textcolor{red}{\XSolidBrush}  & 3.3M & 3.3M \\
    GUIEnv-local & 73K & 9K & / & HTML Text & Point Coordinates & \textcolor{green}{\Checkmark} & \textcolor{red}{\XSolidBrush}  & 700K & 700K \\
    \midrule
    \textbf{\model Collection} & \textbf{173K} & \textbf{104K} & \textbf{7.8K} & \textbf{Diversified Instr.} & \textbf{Refer. Caption + Point Coordinates} & \textcolor{green}{\Checkmark} & \textcolor{green}{\Checkmark} & \textbf{3.9M} & \textbf{11.5M} \\
    \bottomrule
\end{tabular}
    }
\vspace{-0.1in}
\caption{Grounding data of \model compared to existing collections.}
\vspace{-0.2in}
    \label{tab:dataset_comparison} 
\end{table*}

The rapid expansion of graphical user interfaces (GUIs) across web, desktop and mobile platforms has made them indispensable for digital interactions. From completing daily tasks like shopping or booking tickets to complex professional workflows, GUI agents play a critical role in automating these processes.
As illustrated in Figure~\ref{fig:overview}, a typical GUI agent operates in two stages: planning and grounding. In the planning stage, the agent generates action decisions to accomplish the user’s task based on the current screen state as its observation. In the grounding stage, the agent is tasked with locating and interacting with the target element as referred in the instructions provided by planning, thus make actions truly happen in the environment.

While efforts have been put to improve the planning of large multimodal models (LMMs) with CoT~\cite{yao2022react, wei2022chain}, and inference-time scaling~\cite{saha2024system}, effectively grounding GUI elements from language remains a significant challenge. The problem is compounded by the diverse visual layouts across diverse devices, wide variability in planned instructions, and the dynamic nature of task execution in real-world environments, all of which demand robust, adaptable, and efficient solutions.






The basic grounding method involves leveraging HTML or accessibility trees (AXTress, or A11y) to identify the target element. However, feeding long textual contexts of the tree often leads to inefficiencies, hallucination, and biases due to missing information in the tree. The absence of visual input further limits the method's ability to address instructions requiring visual or positional cues. Set-of-Mark~\cite{yang2023som} combines visual and tree tag information. However, its reliance on HTML or AXTrees limits flexibility in diverse environments, as platform standards are inconsistent and, particularly on mobile and desktop, the quality of AXTrees depend largely on app developers' implementation.
Additionally, LMMs struggle to accurately select from numerous tags in images, constraining grounding performance~\cite{xie2024osworld}.
To this end, building a pure-vision solution for GUI agent grounding is crucial.

Training an LMM for GUI instruction grounding is non-trivial.
Existing LMMs are: 1) heavily skewed towards natural images
due to data biases. 2) rarely trained for grounding. While some models are trained with datasets like RefCOCO~\cite{refcoco}, these datasets are not aligned with GUI scenarios and are sparsely populated. Recently, some studies~\cite{cheng2024seeclick, gou2024navigating} have leveraged LMMs’ powerful vision and language capabilities, using public mobile- or web-sourced data as (GUI image, instruction, coordinates) tuples to train LMMs as grounding models. Despite their effectiveness, we identify two key limitations in these approaches: \textbf{(1) They overly depend on rigid instruction sources and formats}, mainly HTML or AXTree-based textual elements. This lack of diversity hinders their robustness in adapting to the flexible and heterogeneous instructions generated by task planners.
\textbf{(2) They overlook the dynamic contextual information during task performing}, such as the action history, which can provide valuable references for more accurate element grounding.

In this paper, we introduce \model, a robust LMM designed specifically for GUI grounding. \model is built upon Aria~\cite{li2024aria}, the state-of-the-art multimodal MoE model with 3.9B activated parameters. \model adopts a pure-vision approach, avoiding reliance on AXTree-like inputs while achieving superior grounding accuracy across diverse tasks and platforms.

By addressing the core limitations of existing methods, we propose two key contributions in \model. For the challenge of rigid instructions, we design a large-scale, diverse data synthesis pipeline from our Common Crawl collection and public available data. This pipeline first leverages strong LMMs to generate detailed and accurate element captions and then utilizes an LLM to create diverse, human-like instructions that align with potential interactions based on these captions. We further incorporate the high-quality captions as additional supervision during training, enabling the model to better associate diverse instructions with their corresponding elements.
For the challenge of ignoring dynamic contexts, we further leverage textual or text-image interleaved action history from trajectory data for training. This equips \model with robust grounding capabilities, enabling it to perform effectively in dynamic, multi-step real-world task scenarios.

To summarize, our contributions are:
\vspace{-0.15in}
\begin{itemize}[leftmargin=*]
    \item We propose a novel approach to address the challenge of rigid instructions with a scalable, data-centric pipeline. It generates high-quality and diverse (element caption, instruction) samples
    from Common Crawl and publicly available data, enabling \model to generalize effectively across diverse instructions in different environments.
    
    \vspace{-0.1in}
    
    \item \model introduces innovative designs for incorporating dynamic action history in textual or interleaved text-image formats. The improvements allow \model to ground elements more effectively in dynamic, multi-step task scenarios, especially under zero-shot settings.  

\vspace{-0.1in}

    \item We conduct comprehensive evaluations on extensive benchmarks including both offline and online agent tasks, showcasing \model's state-of-the-art performance. Notably, \model achieves higher grounding accuracy and task success rates compared to both vision-only and AXTree-reliant baselines. 
\end{itemize}
\vspace{-0.1in}

\vspace{-0.1in}
\section{Method}
\begin{figure*}[t]
    \centering
    \includegraphics[trim=3mm 0 3mm 0, clip, width=1\linewidth]{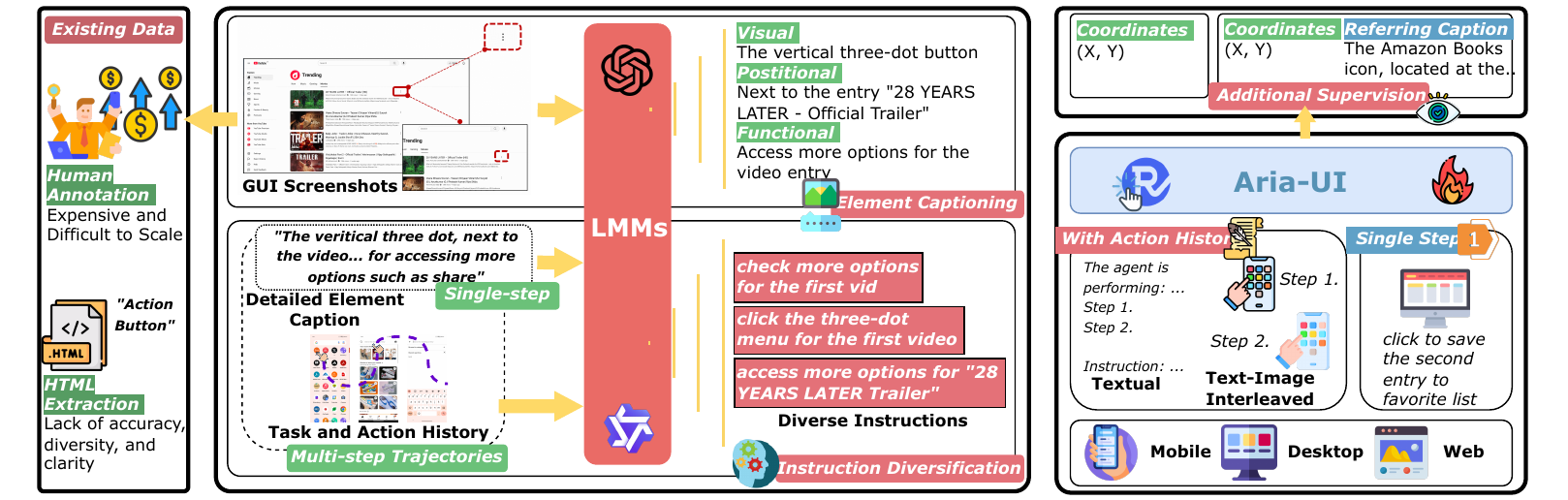}
    \vspace{-0.2in}
    \caption{The overall data and training pipeline for \model.}
    \label{fig:framework}
    \vspace{-0.2in}
\end{figure*}



Aria-UI is designed to seamlessly integrate into the latest general-purpose multimodal GUI agent framework \cite{seeact, xie2024osworld, koh2024visualwebarena, rawles2024androidworld}, serving as a robust grounding model. 
We outline a solution to the challenges from a scalable, data-centric approach, as shown in Figure~\ref{fig:framework}. In Section~\ref{sec:data_pipeline}, we detail the synthesizing of diverse grounding data. Section~\ref{sec:context_data} discusses building grounding samples with task context for dynamic scenarios, and Section~\ref{sec:arch} explains \model’s training details.



\vspace{-0.1in}
\subsection{Large-scale Diverse GUI Data Synthesizing}


As summarized in Table~\ref{tab:dataset_comparison}, several existing methods have collected diverse corpus for GUI grounding. However, these corpora fail to effectively address GUI grounding for LMMs. They are either not open-source, too small, or lack coverage of all the major platforms. Moreover, they rely on rigid instruction sources and formats, from HTML extraction or specifically formatted referring caption. Additionally, they overlook the importance of the contextual information for grounding during dynamic task performing. We present how to solve these challenges by a data-centric approach with diverse data scaling from multiple platforms and context-aware data extension with trajectories.
\vspace{-0.1in}
\subsubsection{Diverse Data Scaling from Multiple Platforms}
\label{sec:data_pipeline}

We propose a two-stage pipeline to transform raw samples into high-quality and diverse element instructions for grounding training. At the first stage, we utilize a strong LMM (GPT-4o or Qwen2-VL-72B~\cite{wang2024qwen2}) that takes element screenshots and text extracted from HTML as input for accurate and detailed element descriptions. To enhance accuracy and reduce hallucination, the model perceives two screenshots: (1) an isolated image of the element and (2) a zoomed-in view, where the element is highlighted with a red bounding box. Additionally, the HTML text and the screen position of the element are provided for reference. The model is then prompted to generate a detailed caption of the element, including its visual properties, functionality, positional relationships, and any other distinctive attributes. In the second stage, we utilize an LLM to generate natural language instructions that correspond to potential interactions with the elements, based on their detailed captions. For instance, for the caption \textit{"The "subscribe" button, colored in bright red with white text and a bell icon, is positioned in the upper-right section of ChefMaria's cooking channel header, showing "2.3M" subscribers” underneath,"} the synthesized instruction could be \textit{"subscribe to ChefMaria's channel."} To ensure diversity and expand the data volume, we produce three instructions for each element.

We apply our pipeline to three key GUI environments: web, desktop, and mobile, each with distinct challenges and characteristics.

\paratitle{Web.}
Web data, with its diversity and dynamic rendering, is ideal for expanding GUI grounding datasets with varied element samples in size, type, and resolution. We leverage the latest collection of Common Crawl for data collection. We build a rigorous data curation and filtering pipeline to produce high-quality samples. We first filter out harmful webpages using fastText~\cite{fasttext}. Subsequently, we identify and select interactive elements by checking the HTML attributes. Considering that LMMs have acquired fundamental OCR skills during pretraining, we prioritize graphical elements over text-based elements. To reflect real-world grounding tasks in complex, element-rich environments, we heuristically retain webpages containing more than 20 valid elements. We use Playwright to render these webpages at 1920$\times$1080 and 2440$\times$1600 resolutions to accommodate common resolution requirements. 
We gather a diverse set of 173K webpages containing 2M elements through the procedure. With the data pipeline, we build detailed caption and instructions for elements, result in 6M high-quality and diverse instruction samples.

\paratitle{Desktop.}
Since desktop environment is less scalable and human annotation costs high, desktop data has remained scarce. OmniACT~\cite{kapoor2025omniact} manually annotated 7.3K instruction-grounding pairs. However, creating an automated data scaling pipeline for desktop remains a challenge. To mitigate the research gap, we develop a traverse agent powered by an LMM to explore an OS environment for data collecting. We build the agent on an Ubuntu Desktop with Gemini 1.5 Flash. Leveraging the accessibility tree, the agent selects the next element to click in each screen state, aiming to reach previously unexplored screens. The implementation details can be found in Appendix~\ref{app:traverse}. We collect all screenshots and the corresponding A11y to parse all elements. Using this automated pipeline, we collected 88K unique elements across 7.8K screenshots tailored for desktop environment. We then utilize the data pipeline to extend the samples to 264K by generating diverse instructions.

\vspace{-0.02in}
\paratitle{Mobile.}
Since automated GUI agents for mobile environments were explored earlier, a substantial amount of open-source data has been accumulated for mobile environment. Currently, the largest-scale grounding dataset for mobile is AMEX~\cite{chai2024amex}, which provides 104K screenshots and 1.6M elements. While AMEX provides a large-scale dataset, it has only 712K elements with basic textual descriptions extracted from accessibility tags, and merely 3K elements are paired with human-like instructions. To address this gap, we regenerate high-quality caption and instruction samples with the data pipeline for AMEX, improving the training effectiveness.

\vspace{-0.02in}
To further expand our grounding corpus and introduce more diverse sources for GUI images and instructions, we incorporate the following public datasets: 3M Web and 273K mobile elements from SeeClick training data~\cite{cheng2024seeclick, li2020widget, li2020mapping}, 15K mobile elements from~\cite{bai2021uibert}, 748k Web elements from GUICourse~\cite{chen2024guicourse}, 131K desktop elements from OmniAct~\cite{kapoor2025omniact}, and 693K Web and mobile elements from AutoGUI~\cite{li2025autogui}. We summarize the details of the datasets in Table~\ref{tab:single_data}.

\vspace{-0.1in}
\subsubsection{Context-aware Data Extension from Trajectories}
\vspace{-0.05in}
\label{sec:context_data}
Accurately and efficiently performing grounding tasks within the dynamic context of real-world environments is a crucial capability for GUI agents. 
Despite its importance, existing approaches largely focus on grounding tasks under a single-step setting, where LMMs are trained to infer grounding results based only on the current state and instruction. Such approaches overlook the dynamic nature of GUI grounding and the critical role of context in real-world scenarios. For example, after executing a \textit{TYPE} action, the next grounding step is likely associated with an \textit{ENTER} or \textit{SUBMIT} button. Similarly, in multi-step tasks that involve navigating through a multi-layered menu to locate a target entry, there is a strong contextual relationship between consecutive grounding actions. Leveraging such contextual information enriches the grounding context and aids the model in avoiding bias, thereby enhancing grounding performance.


We utilize publicly available agent trajectories to simulate grounding tasks with contexts. We focus on constructing two types of contextual setups: (1) textual action history and (2) text-image-interleaved history. The text-based setup incorporates the ultimate task along with prior action histories, and the text-image-interleaved setup extends this by including $N$ historical screen images, providing richer contextual cues and training the model to understand multimodal interaction history. Notably, most trajectory data only includes basic sequential information, such as the click coordinates, thus lacks comprehensive stepwise instruction semantics. To address this, we augment all grounding steps within the trajectory data using the proposed data pipeline to generate detailed stepwise instructions. For non-grounding actions, we encode them (e.g., SWIPE and TYPE) using rules for natural language formats. For the interleaved setting, we collect data with image number $N=[1,2,3]$, and for the text-based setting, we input all historical actions in text. Finally we collect 992K samples with the trajectories from GUI-Odyssey~\cite{guiodyssey}, Android in the Zoo~\cite{aitz}, Android Control~\cite{ac}, Android in the Wild~\cite{aitw} and AMEX~\cite{chai2024amex}. The details are presented in Table~\ref{tab:traj_data}.

\vspace{-0.1in}
\subsection{Model Architecture}
\vspace{-0.05in}
\label{sec:arch}
\begin{table*}[htb]
\centering
\resizebox{0.8\linewidth}{!}{
\begin{tabular}{lccccccr}
\toprule
 \multirow{2}{*}{Method} & \multicolumn{2}{c}{Mobile} & \multicolumn{2}{c}{Desktop} & \multicolumn{2}{c}{Web} & \multirow{2}{*}{Avg.} \\
\cmidrule(lr){2-3} \cmidrule(lr){4-5} \cmidrule(lr){6-7}
 & Text & Icon/Widget & Text & Icon/Widget & Text & Icon/Widget & \\
\midrule
 GPT-4 & 22.6 & 24.5 & 20.2 & 11.8 & 9.2 & 8.8 & 16.7 \\
 GPT-4o & 20.2 & 24.9 & 21.1 & 23.6 & 12.2 & 7.8 & 18.1 \\
 CogAgent & 67.0 & 24.0 & 74.2 & 20.0 & 70.4 & 28.6 & 49.6 \\
 SeeClick & 78.0 & 52.0 & 72.2 & 30.0 & 55.7 & 32.5 & 55.8 \\
 Qwen2-VL & 75.5 & 60.7 & 76.3 & 54.3 & 35.2 & 25.7 & 55.3 \\
 UGround & 82.8 & 60.3 & 82.5 & 63.6 & 80.4 & 70.4 & 74.1 \\
\midrule
  \textbf{\model} & \textbf{92.3} & \textbf{73.8} & \textbf{93.3} & \textbf{64.3} & \textbf{86.5} & \textbf{76.2} & \textbf{82.4} \\

\bottomrule
\end{tabular}
}
\vspace{-0.1in}
\caption{Results on ScreenSpot. We report element accuracy and the micro average results.}
\vspace{-0.2in}
\label{tab:ss}
\end{table*}

We build \model with the state-of-the-art multimodal MoE model, Aria~\cite{li2024aria}. We leverage two strengths from Aria for GUI agents: 1) Aria is multimodal-native, built for better understanding of complex and interleaved contexts; 2) with only 3.9B activated parameters, Aria shows even faster inference speed than 7B dense models.
\vspace{-0.1in}
\subsubsection{Ultra Resolution Support}
\vspace{-0.05in}
With the shift from 1080p to 2K resolutions on computers and mobile devices, training grounding LMMs at high resolutions has become essential. Aria originally supports high-resolution images up to 980×980, which we extend to a maximum of 3920×2940 on \model by splitting the image into smaller blocks, significantly increasing the range of image sizes to handle. 
To maintain positional accuracy, we take inspiration from NaViT~\cite{dehghani2024patch} to place padding before resizing for keeping the original screenshot ratio.



\vspace{-0.1in}
\subsection{Training and Inference Paradigm}
\vspace{-0.1in}
We train \model following a two-phase procedure. We first leverage all the single-step grounding data to train the foundation GUI grounding capability of \model. Specifically, \model is tasked with generating grounding answers given the prompt \textit{"Given a GUI image, what are the relative (0-1000) pixel point coordinates for the element corresponding to the following instruction or description: [...]"}. We follow \cite{gou2024navigating} to group all the samples for the same GUI image into a multi-turn conversation format. Then, context-aware data with both text-based and text-and-image-interleaved history settings are fed into the model to further enhance the grounding capability under the dynamic setting. For this phase, we add extra 20\% samples from the single-step data to keep the generic grounding capability and avoid over-fitting. We place more training details in Appendix~\ref{app:training_details}.

During inference, \model outputs the grounded pixels coordinates normalized to $[0, 1000]$. Since \model is also trained with context-aware trajectories, it can take historical agent actions and grounding actions as chat history, formulating a stronger grounding system in dynamic environments.

\vspace{-0.1in}
\section{Experiments}
\vspace{-0.1in}
\begin{table*}[htb]
\centering
\resizebox{0.8\linewidth}{!}{
\begin{tabular}{lcccccc}
\toprule
\multirow{2}{*}{Models} & \multicolumn{2}{c}{\textbf{AndroidControl-Low}} & \multicolumn{2}{c}{\textbf{AndroidControl-High}} & \multicolumn{2}{c}{\textbf{GUI-Odyssey}} \\
& Grounding & Task SR  & Grounding  & Task SR & Grounding & Task SR \\
\midrule
\multicolumn{7}{c}{\textit{Zero-shot}} \\
\midrule
GPT-4o & 16.36 & 5.12 & 10.36 & 2.84 & 19.66 & 0.05 \\
Qwen2-VL & 64.24 & 32.53 & 30.32 & 4.08 & 49.56 & 2.00 \\
SeeClick & 45.55 & 17.72  & 20.17 & 4.29 & 45.19 & 1.45 \\
UGround & - & - & - & - & 50.25 & 2.02 \\
\textbf{\model} & \textbf{79.70} & \textbf{54.39} &  \textbf{35.12} & \textbf{5.95} & \textbf{64.81} & \textbf{5.28} \\
\midrule
\multicolumn{7}{c}{\textit{W. Training Set}} \\
\midrule
UGround & 74.28 & 46.85  & 37.98 & 9.15 & - & - \\
\textbf{\model} & 85.71 & 66.30  & 41.78 & 9.97 & 84.57 & 31.87 \\
\textbf{\modelth} & \textbf{87.69} & \textbf{67.33}  & \textbf{43.16} & \textbf{10.17} & 86.75 & 36.47 \\
\textbf{\modelih} & 87.20 & 67.26  & 42.97 & 10.10 & \textbf{87.02} & \textbf{37.30} \\
\bottomrule
\end{tabular}
}
\vspace{-0.1in}
\caption{Results for offline mobile agent evaluation. We report element accuracy for grounding and the task success rate. For AndroidControl-High, GPT-4o serves as the planner to generate stepwise instructions for all methods.}
\label{tab:mobile_offline}
\vspace{-0.1in}
\end{table*}

We testify the performances of \model via extensive experiments including single-step grounding, grounding under offline agent trajectories and grounding in dynamic online agent environments.




\subsection{GUI Grounding Evaluation}
We first examine \model's foundational GUI grounding capabilities on ScreenSpot~\cite{cheng2024seeclick}. The benchmark compasses six subsets spanning over two types of elements and three major platforms. Each test entry provides a unique GUI image and a human-annotated instruction for locating a specific element. The typical resolution for mobile and web subsets is 2k, and for desktop samples it is 540p. We include the state-of-the-art UGround~\cite{gou2024navigating}, with previous grounding models SeeClick~\cite{cheng2024seeclick} and CogAgent~\cite{hong2024cogagent} as baselines. We also include generic LMMs -- GPT-4, GPT-4o and Qwen2-VL~\cite{wang2024qwen2}.

From the results in Table~\ref{tab:ss}, \model achieves the highest average accuracy (82.4\%) across all subsets, demonstrating its superior grounding performance. \model achieves a significant margin over the state-of-the-art UGround, particularly excelling in tasks for textual elements. The results showcase \model's robustness and generalizability across diverse platforms and element types.

\subsection{Offline Agent Evaluation}
\begin{table*}[htb]
\centering
\resizebox{0.8\linewidth}{!}{
\begin{tabular}{lllrrrr}
\toprule
Input & Planner & Grounding & Cross-Task & Cross-Website & Cross-Domain & Avg. \\
\midrule
\multirow{2}{*}{Image + HTML Tree} & GPT-4 & Choice & 46.4 & 38.0 & 42.4 & 42.3 \\
 & GPT-4 & SoM & 29.6 & 20.1 & 27.0 & 25.6 \\
\midrule
\multirow{5}{*}{Image} & GPT-4 & SeeClick & 29.6 & 28.5 & 30.7 & 29.6 \\
& GPT-4 & UGround & 45.1 & 44.7 & 44.6 & 44.8 \\
& GPT-4 & OmniParser & 42.4 & 41.0 & 45.4 & 42.9 \\
\cmidrule{2-7}
& GPT-4o & SeeClick & 32.1 & 33.1 & 33.5 & 32.9 \\
& GPT-4o & UGround & 47.7 & 46.0 & 46.6 & 46.8 \\
& GPT-4o & \textbf{\model} & \textbf{56.1} & \textbf{57.0} & \textbf{59.5} & \textbf{57.5} \\
& GPT-4o & \textbf{\model}$_{TH}$ & \textbf{57.6} & \textbf{58.0} & \textbf{61.2} & \textbf{58.9} \\
& GPT-4o & \textbf{\model}$_{IH}$ & \textbf{57.6} & \textbf{57.7} & \textbf{61.4} & \textbf{58.9} \\
\bottomrule
\end{tabular}
}
\vspace{-0.1in}
\caption{Results on Multimodal-Mind2Web, with grounding element accuracy reported. None of the methods adopted the training split, therefore we exhibit a fully zero-shot out-of-distribution evaluation.}
\vspace{-0.25in}
\label{tab:mmm2w}
\end{table*}

\begin{table*}[htb]
\centering
\resizebox{0.8\linewidth}{!}{
\begin{tabular}{lllcc}
\toprule
Input & Planner & Grounding & AndroidWorld & MobileMiniWob++ \\
\midrule
\multirow{2}{*}{AXTree} & GPT-4-Turbo & Choice & 30.6 & 59.7 \\
& Gemini 1.5 Pro & Choice & 19.4 & 57.4 \\
\midrule
\multirow{2}{*}{Image + AXTree} & GPT-4-Turbo & SoM & 25.4 & 67.7 \\
& Gemini 1.5 Pro & SoM & 22.8 & 40.3 \\
\midrule
\multirow{3}{*}{Image} & GPT-4-Turbo & UGround & 31.0 & - \\
& GPT-4o & UGround & 32.8 & 48.4 \\
& GPT-4o & \textbf{\model} & \textbf{39.7} & \textbf{60.4}  \\
& GPT-4o & \textbf{\modelth} & \textbf{44.8} & {-}  \\
\bottomrule
\end{tabular}
}
\vspace{-0.1in}
\caption{Task success rate results for online mobile and Web agents on AndroidWorld and MobileMiniWob++.}
\vspace{-0.1in}
\label{tab:aw}
\end{table*}

\begin{table*}[htb]
\centering
\resizebox{0.9\linewidth}{!}{
\begin{tabular}{lrrrrrrrrrrr}
\toprule
Models & OS & Calc & Impress & Writer & VLC & Thunderbird & Chrome & VSC & GIMP & Multi & Avg. \\
\midrule
GPT-4o + SoM & 20.83 & 0.00 & 6.77 & 4.35 & 6.53 & 0.00 & 4.35 & 4.35 & 0.00 & 3.60 & 4.59 \\
CogAgent + SoM & 4.17 & 2.17 & 0.00 & 4.34 & 6.53 & 0.00 & 2.17 & 0.00 & 0.00 & 0.00 & 0.99 \\
GPT-4o + A11y & 41.67 & 4.26 & 6.81 & 8.70 & 9.50 & 6.67 & 15.22 & 30.43 & 0.00 & 7.46 & 11.21 \\
\midrule
CogAgent & 4.17 & 2.17 & 0.00 & 4.35 & 6.53 & 0.00 & 2.17 & 0.00 & 0.00 & 0.10 & 1.11 \\
GPT-4o & 8.33 & 0.00 & 6.77 & 4.35 & 16.10 & 0.00 & 4.35 & 4.35 & 3.85 & 5.58 & 5.03 \\
GPT-4o + \textbf{\modelth} & \textbf{25.00} & \textbf{4.26} & \textbf{15.32} & \textbf{8.70} & \textbf{30.06} & \textbf{26.67} & \textbf{23.80} & \textbf{21.74} & \textbf{19.23} & \textbf{8.55} & \textbf{15.15}\\
\bottomrule
\end{tabular}
}
\vspace{-0.1in}
\caption{OSWorld results. The top part denotes methods with both accessibility tree (A11y) and screenshot input, while the bottom part is for pure-vision methods that rely only on screenshots.}
\vspace{-0.2in}
\label{tab:os}
\end{table*}

\paratitle{Mobile Agents.}
We further testify how \model performs under an offline dynamic setting, where the model is required to provide grounding coordinates in agent task trajectories. We employ AndroidControl-Low~\cite{ac}, GUI-Odyssey~\cite{guiodyssey} and AndroidControl-High, the first two has human-annotated or generated stepwise instruction, while the last one only provides the user task, and needs an additional planner for stepwise instructions. We follow~\cite{ac, gou2024navigating} to utilize GPT-4o as the planner. We report element accuracy and the task success rate in Table~\ref{tab:mobile_offline}. Specifically, we evaluate \model and the baselines on both zero-shot and training split-included settings. As we evaluate \model with agent trajectories, we extend the model with two variants: \modelth and \modelih, for textual action history input and text-image interleaved history input, separately. We choose $N=1$ for \modelih to include additional one GUI image from history during inference. For \modelth, we always input the full action history.

The results demonstrate the superior performance of \model across different evaluation settings and metrics. Specifically, \model and its variants consistently outperform existing baselines, with \modelth achieving peak performance of grounding accuracy and task success rate on AndroidControl, and \modelih achieving the best performances on GUI-Odyssey. Empirically, we found that the incorporation of historical actions, whether in text-only (${TH}$) or text-image interleaved (${IH}$) format, provides crucial context for accurate element grounding and task completion. In particular, we observe that the textual action history (\modelth) strikes an effective balance between efficiency and performance compared to both the base model and \modelih. 


In summary, the significant performance gap between \model and existing approaches like SeeClick and UGround underscores the effectiveness of our proposed model in understanding and executing mobile interface interactions.

\vspace{-0.05in}
\paratitle{Web Agents.}
We evaluate how \model and its variants perform on multimodal Web agent tasks with the Multimodal-Mind2Web~\cite{deng2024mind2web} benchmark. The original training split is not included by \model and the baselines during the training stage, thus we form a fully zero-shot out-of-distribution scenario. Three subsets, cross-task, cross-website and cross-domain are employed for a comprehensive evaluation.

Shown in Table~\ref{tab:mmm2w}, \model and its variants significantly outperform all baselines across the three subsets, achieving an average accuracy of 57.5\% for the base model and 58.9\% for \modelth and \modelih. Notably, \modelih demonstrates the strongest performance in the cross-website and cross-domain subsets, showcasing its robust ability to leverage historical multimodal context. The improvements over previous models, including UGround and SeeClick, underscore \model's effectiveness in handling zero-shot grounding tasks on diverse and unseen web interfaces.

\vspace{-0.1in}
\subsection{Online Agent Evaluation}
\begin{table*}[htb]
\centering
\resizebox{0.8\linewidth}{!}{
\begin{tabular}{lccccccr}
\toprule
 \multirow{2}{*}{Method} & \multicolumn{2}{c}{Mobile} & \multicolumn{2}{c}{Desktop} & \multicolumn{2}{c}{Web} & \multirow{2}{*}{Avg.} \\
\cmidrule(lr){2-3} \cmidrule(lr){4-5} \cmidrule(lr){6-7}
 & Text & Icon/Widget & Text & Icon/Widget & Text & Icon/Widget & \\
\midrule
{\model} & {92.3} & {73.8} & {93.3} & {64.3} & {86.5} & {76.2} & {82.4} \\
(-) Ultra Resolution & {87.5} & {61.1} & {70.6} & {40.0} & {53.5} & {40.3} & {61.1} \\
(+) Visual CoT Prompting & {93.8} & {59.8} & {80.4} & {51.4} & {73.0} & {57.8} & {71.4} \\
(-) \model Data & {89.0} & {60.7} & {78.3} & {34.3} & {79.6} & {52.9} & {68.7} \\
(-) Diversified Instruction & {88.3} & {67.2} & {83.0} & {57.1} & {82.2} & {63.1} & {74.9} \\
(-) Refer. as Supervision & 92.7 & 69.0 & 81.4 & 54.3 & 85.2 & 70.0 & 77.5 \\

\bottomrule
\end{tabular}
}
\vspace{-0.1in}
\caption{Ablation study results on ScreenSpot.}
\vspace{-0.2in}
\label{tab:ab}
\end{table*}
\paratitle{Mobile and Web.}
We use AndroidWorld~\cite{rawles2024androidworld} for online mobile agent evaluation in an Android emulator environment. The evaluation is fully based on success of the task by checking the system state of the virtual device. We also include the MobileMiniWob++ task collection provided by AndroidWorld, which adpats the Web agent environment MiniWob++~\cite{liu2018reinforcement} to AndroidEnv~\cite{toyama2021androidenv}, the same environment as AndroidWorld. We evalute \model with the strongest baseline, UGround under the same M3A agent framework, compared with SoM and Choice methods that require AXTree input. We report task success rate, the most important metric for real agents in Table~\ref{tab:aw}. Our observations are:
\vspace{-0.1in}
\begin{itemize}[leftmargin=*]
    \item In AndroidWorld, our approach achieves the best performance to date, with a task success rate of 44.8\%, achieved by \modelth. This surpasses the previous state-of-the-art method, UGround, as well as non-pure vision methods such as SoM and Choice, which rely heavily on AXTree input. The results highlight \model's superior ability to handle diverse element instructions in real-world settings, demonstrating its robustness and adaptability for pure-vision GUI agents. 
    \vspace{-0.1in}
    \item On MobileMiniWob++, \model outperforms UGround, and choice-based methods. Due to the simplicity of MiniWob++ layouts, GPT-4-Turbo with SoM achieves the highest performance. However, \model still demonstrates the highest scores with pure-vision input.
\end{itemize}

\vspace{-0.1in}
\paratitle{OSWorld.}
We further evaluate \model on the most up-to-date and complex computer use simulator benchmark, OSWorld~\cite{xie2024osworld}. Following the pure-vision agent framework in OSWorld, we place \model as the grounding model to work collaboratively with GPT-4o on the 369 real tasks provided. We compared \model with previous SOTA methods and summarize the task success rate in Table~\ref{tab:os}. With GPT-4o as planner and \modelth as the grounding model, we achieve the highest average task success rate of 15.15\%, outperforming previous methods across all computer-use scenarios in OSWorld. Notably, it excels in tasks like VLC (30.06\%), Chrome (23.80\%), and Impress (15.32\%), highlighting \model's strong performance in diverse, complex GUI tasks.

\vspace{-0.1in}
\subsection{Ablation Study}
\vspace{-0.05in}
\paratitle{Model Components.}
\vspace{-0.1in}
\begin{itemize}[leftmargin=*]
    \item (-) Ultra Resolution. We remove the ultra resolution support for \model.
    \vspace{-0.1in}
    \item (+) Visual CoT Prompting. We use CoT prompting for \model inference, as in Figure~\ref{fig:aria_ui_prompt_single}.
\end{itemize}
\vspace{-0.1in}



\paratitle{Training Data Ablation.}
\vspace{-0.1in}
\begin{itemize}[leftmargin=*]
    \item (-) \model Pipeline Data. We remove the data from our pipeline during training.
    \vspace{-0.1in}
    \item (-) Diversified Instruction. We directly use refer. caption as input and coordinates as output for training, removing the diversified instructions.
    \vspace{-0.1in}
    \item (-) Refer. as Supervision. We use only coordinates for supervision for our pipeline data.
\end{itemize}
    \vspace{-0.1in}
We summarize the ablation results in Table~\ref{tab:ab}. The results highlight the critical role of ultra resolution (Avg. 61.1) and \model data, particularly for Icon/Widget grounding. Removing diversified instruction or refer. as supervision degrades performance across platforms, due to weak alignment between instruction, refer. caption and grounding coordinates. We also found that adding CoT improves text-based tasks on mobile but struggles with others, caused by noise in visual reasoning.

\paratitle{Context-aware Grounding Effect.} We propose two variants—text-only (${TH}$) and text-image interleaved (${IH}$) for \model—to evaluate the context-aware grounding design. We present the ablation results comparing the two variants and the base \model model in Tables~\ref{tab:mobile_offline}, \ref{tab:mmm2w}, and~\ref{tab:aw}, across both offline and online agent settings. Notably, in the dynamic agent environment AndroidWorld, the context-aware model outperforms the base model by a significant margin of 12.8\%. The results demonstrate that incorporating dynamic context substantially enhances the performance of Aria-UI.



\vspace{-0.05in}
\section{Related Work}
\vspace{-0.05in}

\paratitle{Vision-language Grounding with Large Multimodal Models.} Foundational approaches for vision-language grounding, such as~\cite{zou2023seem, liu2023grounding, li2023semanticsam}, integrate CLIP with specialized vision models to tackle language-guided grounding tasks. To address the limitations in complex reasoning scenarios, researchers have begun leveraging LMMs~\cite{llava, dai2023instructblipgeneralpurposevisionlanguagemodels, shao2024visualcot} as a promising direction. Notable works~\cite{peng2023kosmos, pi2023detgpt, wang2024visionllm} train LMMs to respond to fine-grained language instructions by grounding them in specific visual regions, while general-purpose models~\cite{bai2023qwen, li2024aria} incorporate grounding as a core function during training. Additionally, significant advances in spatial information processing~\cite{zhang2023llava-g, chen2023shikra, zhang2023gpt4roi, you2023ferret, zhang2024ferret2} have enhanced regional visual comprehension capabilities. However, these methods, while effective for natural images, face challenges when applied to GUI screenshots due to insufficient adaptation.

\paratitle{General GUI Agents.}
Automating GUI operations with capable agents has become a trending research area that leverages LMMs. Existing efforts have been put to design autonomous agents for complex task completion on mobile~\cite{rawles2024androidworld,bai2024digirl,li2024appagent,zhang2023appagent,wen2024autodroid,nong2024mobileflow, you2024ferretui, li2024ferretui2}, Web~\cite{koh2024visualwebarena, yao2022webshop, zhou2023webarena, lai2024autowebglm, he2024webvoyager, abuelsaad2024agente, ma2023laser, zhang2024xlam} and desktop~\cite{xie2024osworld, wu2024copilot, gao2023assistgui, zheng2023synapse, zhang2024ufo, niu2024screenagent} environments. These methods initially relied on HTML or AXTrees for element grounding to perform actions. Recently, several notable studies~\cite{cheng2024seeclick, gou2024navigating} have proposed developing pure vision-based GUI grounding models with LMMs. However, due to their lack of instruction diversity and insufficient consideration of dynamic context, these approaches have delivered sub-optimal performances.

\vspace{-0.05in}
\section{Conclusion}
\vspace{-0.05in}
In this paper, we introduced \model, a robust LMM for GUI grounding across diverse environments. We designed a two-stage data pipeline for high-quality and diverse GUI grounding data from multiple platforms. We further incorporated dynamic action history as effective cues for stronger grounding capabilities in real-world environments.
As a scalable and data-centric method, \model outperforms existing methods on all evaluated benchmarks, with both offline and online agent tasks. The model demonstrates strong zero-shot generalization across platforms, establishing \model as a powerful solution for universal GUI grounding.

\clearpage
\section{Limitations}
While \model demonstrates strong performance in grounding target elements based on instructions provided by a planner, it currently lacks the ability to autonomously perform both planning and grounding for a given task. This reliance on the planner model introduces a dependency on the quality and effectiveness of the planner, which can affect overall performance for complex tasks. Additionally, \model’s training does not yet incorporate error correction for planner-generated instructions, limiting its ability to correct mistakes made by the planner during dynamic tasks. Future work will focus on enabling \model to perform integrated planning and grounding, as well as enhancing its ability to handle and correct planner errors in real-time.

\bibliography{custom}

\clearpage
\appendix

\section{Details on \model Datasets}
\subsection{Comprehensive Dataset Statistics}
We summarize the statistical details of the single-step and context-aware grounding datasets in Table~\ref{tab:single_data} and Table~\ref{tab:traj_data}. The key points are:
\vspace{-0.1in}
\begin{enumerate}[leftmargin=*]
    \item Within each platform, the collection of \model dataset possesses the largest sample size, highlighting the scale-up capacity of our automated data pipeline.
    \vspace{-0.1in}
    \item Our collection adopts high-quality diversified instructions for the input text, while other publicly available datasets use plain tags from the tree information, or small-scale human annotations.
    \vspace{-0.1in}
    \item We use diverse trajectory data for context-aware training, with the average steps spanning from 5.5 to 15.4. For the text-based action history setting, we treat all previous actions for a specific grounding step as the context. For the text-image-interleaved setting, we adopt a window size of $N=[1,2,3]$. We then scale up the available training samples to nearly 1M.
\end{enumerate}

\begin{table*}[]
\centering
\resizebox{\linewidth}{!}{
\begin{tabular}{l l l l l}
\toprule
\textbf{Data Collection} & \textbf{Platform} & \textbf{Input Text} & \textbf{\#Screenshots} & \textbf{\#Samples} \\ \midrule
SeeClick~\cite{cheng2024seeclick} & Web & HTML Text & 270K & \textcolor{blue}{3.3M} \\
Widget Captioning~\cite{li2020widget} & Mobile & Instruction & 14.4K & 101K \\
RicoSCA~\cite{li2020mapping} & Mobile & Instruction & 18.1K & \textcolor{red}{173K} \\
UIBert~\cite{bai2021uibert} & Mobile & Refer. Caption & 16.9K & 16.9K \\
GUIEnv~\cite{chen2024guicourse} & Web & HTML Text & 50K & 700K \\
GUIAct~\cite{chen2024guicourse} & Web & Instruction & 13K & 67K \\
OmniACT~\cite{kapoor2025omniact} & Desktop & A11y Text & 7.3K & \textcolor{olive}{131K}\\
AutoGUI~\cite{li2025autogui} & Web \& Mobile & Instruction & 693K & 693K\\
\midrule
\model Web & Web & Diversified Instr. & 173K & \textcolor{blue}{6.4M}\\
\model Mobile (from AMEX~\cite{chai2024amex}) & Mobile & Diversified Instr. & 104K & \textcolor{red}{4.8M} \\
\model Desktop & Desktop & Diversified Instr. & 7.8K & \textcolor{olive}{264K} \\
\midrule
\textbf{Total} & & & \textbf{1.37M} & \textbf{16.6M} \\ \bottomrule
\end{tabular}
}
\caption{Statistics information for \model single-step grounding datasets. The collections with largest size of samples for each platform are highlighted in blue, red and olive, separately. }
\label{tab:single_data}
\end{table*}

\begin{table*}[]
\centering
\resizebox{0.8\linewidth}{!}{
\begin{tabular}{l l l l}
\toprule
\textbf{Data Collection}  & \textbf{\#Avg. Steps} & \textbf{\#Trajectories} & \textbf{\#Samples} \\ \midrule
AitW~\cite{aitw} & 9.67 & 24.5K & 473K \\
AitZ~\cite{aitz} & 7.5 & 2.0K & 26K  \\
AMEX~\cite{chai2024amex} & 12.8 & 3.0K & 68K \\
AndroidControl~\cite{ac} & 5.5 & 13.6K & 156K \\
GUI Odyssey~\cite{guiodyssey} & 15.4 & 7.7K & 269K \\
\midrule
\textbf{Total} & & \textbf{50.8K} & \textbf{992K} \\ \bottomrule
\end{tabular}
}
\caption{Statistics information for \model context-aware grounding datasets.}
\label{tab:traj_data}
\end{table*}

\vspace{-0.1in}
\subsection{Pseudo-code for Desktop Grounding Data Scaling Agent Implementation}
\label{app:traverse}
We present the pseudo-code for OS traverse agent with LMM (Gemini 1.5) guidance in Figure~\ref{fig:pseudo_code}. To summarize, the key ideas for developing the agent are:
\vspace{-0.1in}
\begin{enumerate}[leftmargin=*]
    \item The system is designed as a heuristic depth-first search over the OS environment. A large multimodal model (Gemini 1.5) is employed to prioritize informative UI elements that are more likely to lead to novel system states, while avoiding interactions with exit elements until other options are explored.
    \vspace{-0.1in}
    \item We represent the system state by hashing all intractable elements on the current screen, using these hashes as unique state IDs. Empirically, this approach effectively identifies identical system states and discriminates between similar yet distinct states.
    \vspace{-0.1in}
    \item Operating system hotkeys, such as \textsc{esc} and \textsc{space}, are used to force a transition from the current state to a previous one for backtracking purposes.
\end{enumerate}

\begin{figure*}[]
\begin{blue-box}[label={fig:pseudo_code}]{OS Traverse Agent with LMM Guidance}
\begin{lstlisting}
1. Initialize:
   - stack = [(start_state, 0)]  # Stack holds (state, depth)
   - visited = {}  # Track visited states using hashed entrances
   - memory = {}  # Store state transitions

2. While stack is not empty:
   - (current_state, depth) = stack.pop()
   - current_hash = hash(extract_entrances(current_state))  # Hash entrances to represent state
   - If current_hash in visited or depth > max_depth: continue
   - visited.add(current_hash)

    2.1. Extract entrances:
        - entrances = extract_entrances(current_state)  # From accessibility tree

    2.2. Rank entrances using LMM:
        - selected_entrances = LMM.rank_and_select(entrances, memory)

    2.3. For each entrance in selected_entrances:
        - next_state = simulate_interaction(current_state, entrance)
        - next_hash = hash(extract_entrances(next_state))
        - If next_hash not in visited: 
            - stack.push((next_state, depth + 1))

    2.4. Update memory:
        - memory[current_state] = selected_entrances

    2.5. Log transitions:
        - Record (next_state = current_state) for debugging or visualization.
\end{lstlisting}
\end{blue-box}
\end{figure*}

\subsection{Discussion on the Overlap between OSWorld Tasks and the Desktop Data}
As discussion in Section~\ref{sec:data_pipeline}, we traverse and scale-up desktop GUI data from general OS and software functionality interfaces on Ubuntu. For clarification, in the process, we do not target any specific downstream tasks in OSWorld, and we strictly do not load any task-specific file or pre-defined system states from OSWorld configurations, to prevent data leakage. For example, in Chrome, we traverse settings and general functionality but avoid browsing any specific webpages or configured set of pages that are part of the test examples. Similarly, in Impress and VLC, we collect general interface functionalities without accessing any specific slides or videos used in the test. Further, we only collect general single-app grounding screenshots. But the test tasks in OSWorld are more complex -- 1/3 of them involves multiple opened windows on the screen with specific contents. Finally, our data collection also extends to other parts of the OS (e.g., Ubuntu’s system settings and app store), which are not within the scope of the OSWorld benchmark tasks.

\section{\model Training Details}
\label{app:training_details}
We train \model with 64 NVIDIA H800 GPUs using the Megatron-LM~\cite{shoeybi2019megatron} framework. For the phase-1 single-step grounding training, we use a context length of 4096, which is kept the same as the original Aria-base model. The phase-1 training takes 18 hours with 10K steps. The phase-2 training aims at enabling \model's context-aware grounding ability with trajectory-based samples. In this stage, we extend the model's context length to 8192 to accommodate long trajectory data. It takes 6 hours for the phase-2 training with 2K steps. During training, expert parallelism (EP) was enabled with a factor of 8 to balance workload distribution across GPUs. Global batch size is set to 256, and we use a learning rate of 2e-5 for phase-1 training and 1e-5 for phase-2 training, separately. The minimum learning rate is set to 1e-8. We enable ViT training and set the learning rate of ViT parameters to $1/10$ of the LLM's learning rate for better performance in multimodal multi-task learning~\cite{li2024aria}.
\section{Prompts for Data Augmentation and Model Inference}
In the following boxes, we present comprehensive and detailed prompts we use for \model's data augmentation, training and inference.

\begin{figure*}[t]
\begin{blue-box}[label={fig:caption_prompt}]{System Prompt for Element Captioning}
\begin{lstlisting}
Given two related images:
A context image: One of a 1x3 grid layout from a full UI screenshot, where a specific element is highlighted in a red box
A detail image: The isolated element itself

Task: Provide a comprehensive sentence describing the button by combining:
Visual properties (text, shape, color, icons)
Functionality (what the button does)
Position (both within the screen layout and in relation to nearby elements)
Any distinctive visual characteristics

The description must include ALL these elements, structured naturally in concise and accurate sentences.

Position descriptions should reference both:
Screen quadrant location (e.g., top-left, bottom-right)
Relative position to surrounding elements

Guidelines:
Do not mention the red highlighting box
Keep the description concise but complete
Include all specified properties
Reference surrounding elements for context
Incorporate the provided screen relative position: {relative_position}
Do not mention the position to the cropped image, only to the full screen layout

Satisfy these requirements to receive a reward. Failure to do so will result in a penalty.

Starting with "The [short text/shape/visual feature] [button/icon/menu/image/bar/...]". If the element has no text, use the most prominent icon or shape.
\end{lstlisting}
\end{blue-box}
\end{figure*}

\begin{figure*}[t]
\begin{blue-box}[label={fig:aria_ui_prompt_single}]{System Prompt for \model Single-step Grounding Training}
\begin{lstlisting}
<Input>
<|img|>Given a GUI image, what are the relative (0-1000) pixel point coordinates for the element corresponding to the following instruction: {instruction}

<Output>
```
{coordinates}
```

<CoT Input>
<|img|>Given a GUI image, what are the relative (0-1000) pixel point coordinates for the element corresponding to the following instruction: {instruction}
Think step-by-step, provide referring for the element first and then the grounded point coordinates.

<CoT Output>
```referring
{elem_caption}
```
```grounding
{coordinates}
```
\end{lstlisting}
\end{blue-box}
\end{figure*}

\begin{figure*}[t]
\begin{blue-box}[label={fig:aria_ui_prompt_traj}]{System Prompt for \model Context-aware Grounding Training}
\begin{lstlisting}
The agent is performing the ultimate task: {ultimate_task}.

History of the agent's steps:\n{history_list}.

<Text-Image Interleaved Context>
<|img|>Step {step_idx}. Instruction: {prev_instruction}

<|img|>Step {step_idx}. Given a GUI image, what are the relative (0-1000) pixel point coordinates for the element corresponding to the following instruction or description: {instruction}
\end{lstlisting}
\end{blue-box}
\end{figure*}

\begin{figure*}[t]
\begin{blue-box}[label={fig:instruction_prompt}]{System Prompt for Instruction Diversification}
\begin{lstlisting}
Attention! You know a lot about GUIs on mobile, desktop and web. Given a detailed description of a GUI element, your task is to generate several user-oriented instructions that would require interacting with the GUI element.
For example:
Input: The 'Search Jobs' button, located at the center-right part of the image and just below the search bar, features a magnifying glass icon on a blue background, indicating its function to initiate a job search.
Output: The 'Search Jobs' button is key to starting or updating a job search after users enter or change their criteria. Instructions should focus on interacting with this button directly after inputting search terms or making adjustments.
```
"search for jobs."
"initiate job search."
"retry the job search."
"begin a new search"
```

Include the important identifications of the specific object to interact with. Examples:
Input: The "subscribe" button, colored in bright red with white text and a bell icon, is positioned in the upper-right section of ChefMaria's cooking channel header, showing "2.3M subscribers" underneath.
Output: (your reflection here)
```
"subscribe to ChefMaria's channel"
"click subscribe on ChefMaria's cooking channel"
```

First do a short reflection and give your answers that involve several possible user instructions. Wrap your answers in ``` as in the example. Use \n to separate multiple possible instructions.

{elem_caption}
\end{lstlisting}
\end{blue-box}
\end{figure*}




\end{document}